\documentstyle[twocolumn,prl,aps,epsfig,amsfonts]{revtex}

\begin{document}
\draft
\wideabs
{
\title{ The Dynamics of the Formation of Degenerate Heavy Neutrino Stars}
\author{Neven Bili\'{c}\cite{bilicadd}, Robert J. Lindebaum\cite{rjladd}, Gary B. Tupper and Raoul D. Viollier}
\address{Institute of Theoretical Physics and Astrophysics,
 Department of Physics, University of Cape Town,
 Private Bag, Rondebosch 7701, South Africa; Email: viollier@physci.uct.ac.za
}
\maketitle
\begin{abstract}
Through reformulating the cold, self-gravitating fermion gas as a Bose
condensate by identifying their mutual Thomas-Fermi limits, the 
dissipationless formation of a heavy neutrino star in gravitational collapse 
is numerically demonstrated. Such stars offer an alternative to supermassive 
black holes for the compact dark objects at the centers of galaxies.
\end{abstract}
}

Supermassive neutrino stars, in which gravity
is balanced by the degeneracy pressure of cold fermions, have been 
a subject of speculation for more than three decades \cite{1}. 
Originally, these 
objects were proposed as models for dark matter in galactic halos and clusters
of galaxies, with neutrino masses in the $\sim$ eV range. More recently,
however, 
degenerate superstars composed of weakly interacting fermions in the $\sim$
10 keV range have been suggested as an alternative to the supermassive black 
holes that are purported to exist at the centers of galaxies  \cite{2,3,4,5,6}.
In fact it has been shown \cite{4}
that such degenerate fermion stars 
could explain the whole range of supermassive compact dark objects which 
have been observed so far, with masses 
ranging from 10$^{6}$ to 3 $\times$ 10$^{9}$$M_{\odot}$, merely assuming
that a weakly interacting quasi-stable fermion of mass $m_{f} \simeq$ 15 
keV exists in nature.
                          
As an example, the most massive and violent compact dark object ever
observed  is located at the center of M87 with a mass $M \simeq$ 3.2
$\times$ 10$^{9} M_{\odot}$ \cite{7}. Interpreting this as a
relativistic fermion star at the  Oppenheimer-Volkoff \cite{8} limit
yields the fermion mass $m_{f} \simeq$ 15  keV and a radius $R =$ 4.45
$R_{\rm S} \sim$ 1.5 light-days \cite{3,4}, where $R_{\rm S}$  is
Schwarzschild radius. In this case there is little difference between
the fermion star and black hole scenarios because the last stable
orbit  around a Schwarzschild black hole is at 3$R_{\rm S}$ anyway.

Extrapolating this down to the compact dark object at the center of
our galaxy  \cite{9}, with mass $M \simeq$ 2.6 $\times$ 10$^{6}
M_{\odot}$, which is at the lower limit of  the mass range, and $R \;
\lesssim$ 20 light-days, the same fermion mass gives $R \; \sim$
10$^{4} \; R_{\rm S}$. Consequent upon the shallow potential inside
this fermion  star, the spectrum of radiation emitted by accreting
baryonic matter is cut  off for frequencies larger than 10$^{13}$ Hz
\cite{3,5}, as is observed in the  spectrum of the strong radio source
Sgr A$^{*}$ at the galactic center \cite{10}.  This fermion star is
also consistent \cite{6} with the observed motion of stars  within a
projected distance of 10 to 30 light-days of Sgr A$^{*}$ \cite{9}.

Of course, it is well-known that 15 keV lies squarely in the
cosmologically  forbidden mass range for stable active neutrinos $\nu$
\cite{11}. Sterile neutrinos are  another matter: as shown by Shi and
Fuller \cite{12}, in the presence of an initial  lepton asymmetry of
$\sim$ 10$^{-3}$, a sterile neutrino $\nu_{s}$ of mass  $m_{s} \sim$
10 keV is resonantly produced with near closure density, $\Omega$ =
1. Moreover, the resulting  energy spectrum is not thermal but rather
cut off so as to approximate a cold  degenerate Fermi gas. This model
is constrained by astrophysical bounds on  $\nu_{s} \rightarrow \nu
\gamma$ \cite{13}, however the allowed parameter space includes  $m_{s}
\simeq$ 15 keV contributing $\Omega_{s} \simeq$ 0.3 as favoured by the
BOOMERANG data \cite{14}.

The statics of degenerate fermion stars is well understood, being the
Oppenheimer-Volkoff equation in the relativistic case \cite{8}   or
the Lan\'{e}-Emden equation with polytropic index $n$ = 3/2 in the
nonrelativistic limit \cite{15}. Alternatively, because $R \gg$
1/$m_{f}$, one may understand these as the Thomas-Fermi theory applied
to self- gravitating systems. The extension of the Thomas-Fermi theory
to finite  temperature \cite{23,16} has been used to show that at a
certain critical temperature  weakly interacting massive fermionic
matter undergoes a first-order  gravitational phase transition from a
diffuse to a clustered state, i.e.\ a  nearly degenerate fermion
star. Such studies do not, however, bear on the  crucial dynamical
question of whether the fermion star can form through  gravitational
collapse of density fluctuations in an orthodox cosmological
setting. Indeed, since collisional damping is negligible,  one would
expect  that only a virialized cloud results \cite{11}.

$N$-body simulations of the collisionless Boltzmann or Vlasov equation
evidence  a rather different picture: the collapse is followed by a
series of bounces  with matter expelled at each bounce, leaving behind
a condensed object \cite{17}.  By Liouville's theorem the Vlasov
equation  describes an incompressible fluid in phase-space so that it
respects a form of  the exclusion principle. Hence, these $N$-body
simulations are effectively  fermion simulations. What transpires is
that gravity, being attractive, self- organises the phase-space fluid
into a high-density/momentum core at the  expense of other
low-density/momentum regions as seen in the evolution of the
spherical Vlasov equation \cite{18}.

Much the same behaviour is observed in the formation of mini-boson
stars  through so-called gravitational cooling \cite{19}. Such a
mini-boson star is stable by  balancing uncertainty and gravitational
pressure. A similar  mechanism works in  the presence of a quartic
self-interaction \cite{20} which dominates over  uncertainty pressure
resulting in an equilibrium radius $R \gg$ 1/$m_{b}$  where $m_{b}$ is
the boson mass \cite{21}. Hence we have a universal description of
the physics underlying the formation process: once the collapse
proceeds far  enough (uncertainty, interaction or degeneracy) pressure
results in a bounce,  the outgoing shock wave carrying away the
binding energy. The virial  argument above  is circumvented because
the ejected matter invalidates its assumption that  there is no flow
through the boundary.

%...........
In this letter we verify the above picture for the formation of the
fermion star from a cold nonequilibrium configuration.  The dynamical
Thomas-Fermi theory was given long ago by Bloch for the electron gas
\cite{22}, and amounts to Euler's equations for irrotational flow
together with an  equation of state $P = P(\rho)$. The problem is
that, transcribed to the self-gravitating fermion gas, there is the
Jeans instability, signalled by an imaginary plasma frequency, and
thus short-wavelength shocks must be regulated.  The usual remedy is
to introduce  some small numerical viscosity; however it seems
imprudent to draw conclusions  based on introducing dissipation into
what is fundamentally a dissipationless  process. Here we take
another, literally conservative approach.

In the Newtonian limit a self-interacting boson star is governed by
the Gross-Pitaevskii-like equations
%............. 
%EQ 1
\begin{mathletters}
\begin{equation}
i \frac{\partial \psi}{\partial t} \; = \; \left[ - \frac{\Delta}{2
 m_{b}} + V \left( | \psi|^{2} \right) + m_{b} \varphi \right] \; \psi
\end{equation}
\begin{equation}
\Delta \varphi \; = \; 4 \pi G  \; | \psi |^{2} \; ,
\end{equation}
\end{mathletters}  
where for convenience we have absorbed the boson mass $m_{b}$ in the
field.  Using the ansatz $\psi = \sqrt{\rho}$  exp  ( - $i m_{b}
\theta$), we arrive at
%EQ 2
\begin{mathletters}
\begin{equation}
\frac{\partial \rho}{\partial t} \; = \; \vec{\nabla} \cdot ( \rho
\vec{\nabla} \theta)
\end{equation}
\begin{equation}
\frac{\partial \theta}{\partial t} \; = \; \frac{(\vec{\nabla}
\theta)^{2}}{2} + \frac{1}{m_{b}} \; V (\rho) + \varphi - \frac{1}{2
m_{b}^{2} \sqrt{\rho}} \; \Delta \sqrt{\rho} \; \; .
\label{eq2b}
\end{equation}
\end{mathletters}

The Thomas-Fermi limit is governed by $m_{b} \gg |\vec{\nabla} \rho|
\; /  \rho$.  Thus neglecting the last term in (\ref{eq2b}), Bloch's equations
are recovered, with $\theta$ being the velocity potential and
$V(\rho)$ given by
%EQ 3
\begin{equation}
V(\rho) \; = \; m_{b} \; \int \; \frac{dP}{\rho} \; .
\label{eq3}
\end{equation}
For a general polytropic equation of state
%EQ 4
\begin{equation}
P (\rho) \; = \; K \; \rho^{1+1/n}
\end{equation}
eq.(\ref{eq3}) yields
%EQ 5
\begin{equation}
V (\rho) \; = \; (n + 1) \; K m_{b} \; \rho^{1/n} \; .
\label{eq5}
\end{equation}
Using the potential (\ref{eq5}) and introducing
%EQ 6
\begin{mathletters}
\begin{equation}
R_{*} \equiv \frac{\Lambda}{m_{b}} \equiv \frac{\left[ (n + 1) K
\right]^{n/2}}{\sqrt{4 \pi G}}
\end{equation}
\begin{equation}
M_{*} \; = \; \frac{R_{*}}{ G}
\end{equation}
\end{mathletters}
as the length and mass scales, respectively, the substitution
%EQ 7
\begin{equation}
\psi \; = \; \frac{\Psi}{\left[ (n+1) K \right]^{n/2}}
\end{equation}
yields the dimensionless equations
%EQ 8
\begin{mathletters}
\begin{equation}
\frac{i}{\Lambda} \; \frac{\partial \Psi}{\partial t} \; = \; \left[ -
\frac{\Delta}{2 \Lambda^{2}} + \varphi + | \Psi |^{2/n} \right] \; \Psi
\label{eq8a}
\end{equation}
\begin{equation}
\Delta \varphi \; = \; | \Psi |^{2} \; .
\end{equation}
\end{mathletters}
The validity of the Newtonian approximation in the static case requires
%EQ 9
\begin{equation}
M/M_{*} \; = \; (4\pi)^{-1}\int \; d^{3}r \; | \Psi |^{2} \ll 1 \; ,
\hspace{.5cm} n < 3  \; .
\end{equation}
For weakly interacting degenerate fermions, the
polytropic index is $n$ = 3/2, and
%%EQ 10
\begin{mathletters}
\begin{equation}
R_{*} \; = \; \left( \frac{9 \pi^{2}}{32 g_{f}^{2}} \right)^{1/4} \;
\frac{m_{\rm Pl}}{m_{f}^{2}} \; = \; 0.2325 \; \left(
\frac{\mbox{keV}}{m_{f}} \right)^{2} \; \sqrt{\frac{2} {g_{f}}} \;
{\rm lyr} \; ,
\end{equation}
\begin{equation}
M_{*} \;  = \; \mbox{1.185} \left( \frac{{\rm keV}}{m_{f}} \right)^{2}
\; \sqrt{ \frac{2}{g_{f}}} \times 10^{11} \; M_{\odot} \; ,
\end{equation}
\end{mathletters}
where $g_{f}$ is the spin degeneracy and $m_{\rm Pl}$ is the Planck
mass.  By construction a large but finite $\Lambda$ allows us to
simulate the  fermionic  problem as a bosonic one through their mutual
Thomas-Fermi limits, while  providing an explicitly energy conserving
way of controlling the shocks.  The basic regulating mechanism is the
``kinetic'' part of (\ref{eq8a}) which penalises gradients of order
$\Lambda$. Of course, $\Lambda$ must be sufficiently large that this
term does not change the static scaling relationship
\begin{equation}
M \; R^{\left( \frac{3-n}{n-1} \right)} \; = \; {\rm const}
\end{equation}
arising from the polytropic equation of state.  Our criterion is that
the ratio of ``kinetic'' and ``pressure'' contributions to the static
energy functional should be small, in  particular for a Gaussian $\Psi
= \alpha \; {\rm exp} \; \left[ - (r/\beta)^{2} \right]$
%EQ 12
\begin{equation}
\frac{T[\Psi]}{P[\Psi]} \; = \; \left( 1 + \frac{1}{n} \right)^{3/2}
\; \frac{3}{2 \; \Lambda^{2} \; \beta^{2} \; \alpha^{2/n}} \; .
\end{equation}
For $n$ = 3/2 this is independent of the size $\beta$ for a given
mass, yielding the weak condition $\Lambda \gg$ 0.97
($M_{*}/M$)$^{1/3}$.

\begin{figure}
\epsfig{file=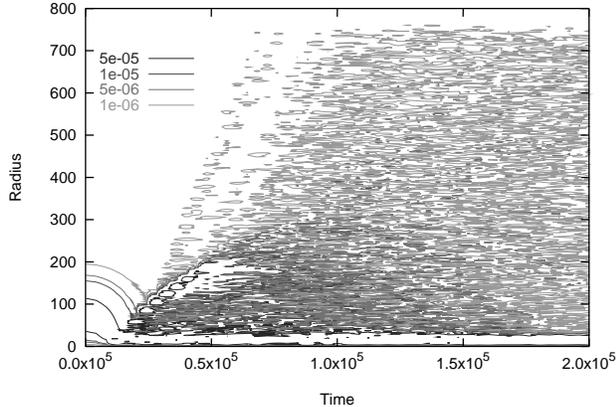,width=5.7cm,angle=270}
\caption{Contour plot for the evolution of $|r \Psi |^{2}$ from the
initial configuration in text.}
\label{fig1}
\end{figure}
\begin{figure}
\epsfig{file=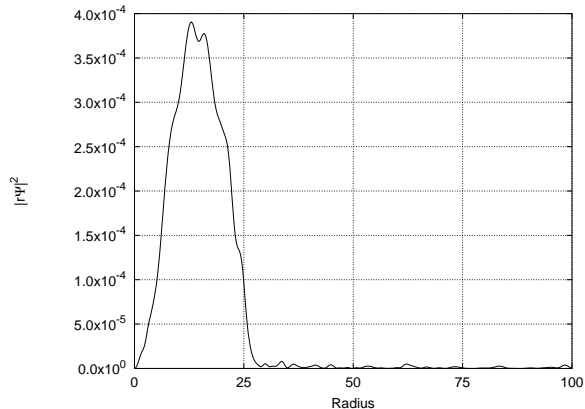,height=5.5cm}
\caption{ $| r \Psi |^{2}$ versus $r$ on the final time slice in
Figure 1.}
\label{fig2}
\end{figure}

In Fig.\ \ref{fig1} we display the evolution of $|r \Psi |^{2}$ for spherical
collapse of a mass $M$ = 0.008 $M_{*}$, initially in the form of a
Gaussian with $\beta$ = 100 = $\Lambda$. The expected features of
bounce and ejection leaving a condensed core are evident. Here we have
implemented a velocity dependent imaginary part to the potential in
the outer layers of the cavity to remove the ejected fermion matter
before it can be artificially reflected by the boundary.  The core
size $r/R_{*}=26$ and mass $m/M_{*}=0.0057$ is commensurate with a
fermion star, however it is far from smooth as evidenced
by the plot of $|r \Psi |^{2}$ on the final time slice in Fig.\ \ref{fig2}. 
This feature may, however, be attributed  to the  relatively short
duration of the simulation.

In summary, using a bosonic representation of the dynamical Thomas-Fermi 
theory for a self-gravitating gas, we have shown that nonrelativistic, 
degenerate and weakly interacting fermionic matter will form supermassive
fermion stars through gravitational collapse accompanied by ejection.
For a fermion mass of $m_{f} \simeq$ 15 keV such a superstar is consistent 
with observations of the compact dark object at the center of our galaxy. A 
similar demonstration for formation near the Oppenheimer-Volkoff limit, and 
the question of cosmology with degenerate dark matter requires a general 
relativistic extension which is under development and will be reported 
elsewhere.

%________________________
The authors wish to thank Duncan Elliott %\footnote{Deceased 20 July 2000} 
 for many useful discussions regarding the simulations. 
This research is in part supported by the Foundation of Fundamental Research 
(FFR) grant number PHY99-01241 and the Research Committee of the University of 
Cape Town.
The work of N.B. is supported in part by the Ministry of Science and 
Technology of the Republic of Croatia under Contract No.\ 00980102.

%\newpage
%\begin{figure}
%\caption{Contour plot for the evolution of $|r \Psi |^{2}$
%           from the initial configuration in text.}
%\label{fig1}
%\end{figure}
%\begin{figure}
%\caption{ $| r \Psi |^{2}$ versus $r$ on the final time slice
%          in Figure 1.} 
%\label{fig2}
%\end{figure}
\end{document}